\documentclass[reprint,pra,groupedaddress,showpacs,preprintnumbers,amsmath,amssymb]{revtex4-1}
\usepackage{graphicx}
\usepackage{dcolumn}
\usepackage{amsmath}
\usepackage{times}
\usepackage{color}
\usepackage{float}
\usepackage[section]{placeins}
\usepackage{physics}
\usepackage{bm}

\newcommand{\Beq}{\begin{equation}}
\newcommand{\Eeq}{\end{equation}}

\renewcommand{\vec}[1]{\bm{#1}}

\newcommand{\BEq}{\begin{eqnarray}}
\newcommand{\EEq}{\end{eqnarray}}
\newcommand{\BsEq}{\begin{subequations}}
\newcommand{\EsEq}{\end{subequations}}
\newenvironment{frcseries}{\fontfamily{frc} \fontsize{9}{11}\selectfont}{}

\begin{document}

\title{Integral equation approach for a hydrogen atom in a strong magnetic field }

\author{B.\ P.\ Carter}
\author{Z.\ Papp}
\affiliation{ Department of Physics and Astronomy,
California State University Long Beach, Long Beach, California, USA }

\date{\today}

\begin{abstract}
The problem of a hydrogen atom in a strong magnetic field is a notorious example of a quantum system that has genuinely 
different asymptotic behaviors in different directions. 
In the direction perpendicular to the magnetic field the motion is quadratically confined, while
in the direction along the field line the motion is a Coulomb-distorted free motion.
In this work, we identify the asymptotically relevant parts of the Hamiltonian and cast the problem into a Lippmann-Schwinger form.
Then, we approximate the asymptotically irrelevant parts by a discrete Hilbert space basis that allows an exact analytic
 evaluation of the relevant Green's operators by continued fractions. The total asymptotic Green's operator is calculated
 by a complex contour integral of subsystem Green's operators. We present a sample of numerical results for a wide range of magnetic field strengths.

\end{abstract}


\maketitle

\section{Introduction}

The problem of a hydrogen atom in a magnetic field goes all the way back to the early history of quantum mechanics. 
The interpretation of how the spectral lines split in a weak magnetic field, the Zeeman effect, or when the magnetic interaction 
dominates over the spin-orbit term, the Paschen-Back effect, provided early support for quantum mechanics. 
In these cases, the Coulomb interaction is dominant over the magnetic interaction, so the latter can be treated as perturbation.

However, when the strength of the magnetic field increases and becomes similar to the strength of the Coulomb field, 
neither of them is small compared to the other and neither of them can be treated as perturbation. 
It was pointed out in  Ref.\ \cite{rau2012topics}  that 
this kind of problem could have been studied as early as 1925, 
but there was no motivation to do so given that only weak magnetic fields were available compared with the internal fields in an atom.
This situation changed in the late 1950s,
when it became evident that the absorption of photons by some semiconductors in magnetic fields then available
depended on the properties of Mott excitons \cite{elliott1960theory}.
Additional motivation to study atoms in very strong magnetic fields was provided in the 1960s by the observation of pulsars,
later identified as neutron stars. 
A rather recent and detailed 
account of the astrophysical aspect of the problem is given in Ref.\ \cite{rau2012topics}. 
 
 Several non-perturbative methods have been developed so far; most of them are adiabatic or variational methods.
The aim of this work is to present a method which is equally applicable for weak and strong magnetic fields.
The difficulty of the problem comes from the fact that the system has genuinely different type of asymptotics in different directions.
In the direction perpendicular to the magnetic field the motion is quadratically confined, 
while along the magnetic field the motion is a Coulomb distorted free motion.
In our approach, we cast the Schr\"odinger equation into a Lippmann-Schwinger 
form and approximate the potential term in a discrete Hilbert space basis.
The corresponding asymptotic Green's operator is calculated by a contour integral. 
In Sec.\ II we introduce the Hamiltonian and set up the Lippmann-Schwinger equation. 
In Sec.\ III we present the solution method.
Then, in Sec.\ IV we present numerical results and then summarize our findings with conclusions.

\section{Magnetic hydrogen atom in Lippmann-Schwinger form}

We treat the  hydrogen atom as  a non-relativistic spin-$1/2$ electron moving 
under the combined influence of a Coulomb center and a homogeneous magnetic field $\mathbf B$ oriented along the $z$-axis.  
We adopt atomic units, such that $\hbar=e=m_{e}=1$. 
The Pauli Hamiltonian is given by
\Beq
\hat{H} = \frac{1}{2m_{e}}  \left( \vec{\sigma} \cdot \left(\hat{\vec{p}}+ \frac{e}{c} \vec{A}\right)   \right)^{2} + U({r}),
\Eeq
with $\vec{A} = (\vec{B}\cross \vec{r})/2$ and $U({r}) = Z/r=Z/\sqrt{\rho^{2}+z^{2}}$ with $Z=-1$.
In cylindrical coordinates, the Hamiltonian takes the form
\Beq
\hat{H} = - \frac{1}{2} \pdv[2]{\rho} -\frac{1}{2} \pdv[2]{z} + \frac{1}{8} {\cal H}^{2} \rho^{2} + \frac{1}{2} {\cal H} (l_{z}+\sigma_{z}) 
+ U(r),
\label{hhh0}
\Eeq
where ${\cal H} = B/B_{a}$ is the dimensionless reduced magnetic field, 
$B_{a} = m_{e}^{2}e^{3}c/\hbar^{3}$ is the atomic unit of magnetic field strength, and $l_{z}$ and $\sigma_{z}$ are the
$z$-components of the orbital and spin angular momenta, respectively.

This system possesses cylindrical symmetry, which results in the conservation of $l_{z}=0,\pm 1, \pm 2, \ldots$ and
$\sigma_{z}= \pm 1$. 
The Hamiltonian in Eq.\ (\ref{hhh0}) is also symmetric with respect to mirroring on the $x-y$ plane, the 
$z \to -z$ parity transformation. The solution should be an eigenstate
 of the  $z$-parity operator  $\hat{{\cal P}}_{z}$, with eigenvalues $\pi_{z} = \pm 1$ for even and odd states, respectively.
So, $\{\hat{H}, \hat{l}_{z}, \hat{\sigma}_{z},\hat{{\cal P}}_{z} \}$ form a complete set of commuting observables and the physically 
admissible  states are their common eigenstates. Since the angular momentum terms in Eq.\ \eqref{hhh0} are trivial, 
we separate them off and 
consider only the eigenvalue problem with the Hamiltonian
\Beq
\hat{H} = - \frac{1}{2} \pdv[2]{\rho} + \frac{1}{8} {\cal H}^{2} \rho^{2}  -\frac{1}{2} \pdv[2]{z} + U(r).
\label{hhh}
\Eeq

We can see from the Hamiltonian in Eq.\ \eqref{hhh} that this system is quadratically confined in the $\rho$ direction and has Coulomb like
asymptotics in the $z$ direction. Therefore we take  the asymptotic Hamiltonian as a sum of Hamiltonians
in the $\rho$ and $z$ directions
\Beq
\hat{H}_{a} = \hat{h}_{\rho} + \hat{h}_{z}.
\label{eq4}
\Eeq
In Eq.\ \eqref{eq4}, we have
\Beq
\hat{h}_{\rho} = \frac{\hat{{p}}^{2}_{\rho}}{2}    +\frac{1}{2} \omega_{h}^{2} \rho^{2},  
\label{Hrho}
\Eeq
with $\omega_{h} = {\cal H}/2$,
and
\Beq
\hat{h}_{z} =  \frac{\hat{{p}}^{2}_{z}}{2} + v(z),
\label{Hz}
\Eeq
where $v(z)$ behaves in the $\abs{z}\to \infty$ limit as $Z/\abs{z}$. For $v(z)$, a convenient choice would be $v(z)=Z/\abs{z}$. 
This potential, however, has an essential singularity at $z=0$. This singularity separates the positive and and negative $z$ parts of the wave
function, forcing the wave function to vanish at $z=0$. 
This rules out positive $z$-parity solutions, which are admissible solutions of the original problem. Therefore we adopt 
\Beq
v(z) =   Z\frac{\erf{(a_{0} z)}}{z},
\label{erfpot}
\Eeq
which has the required Coulomb asymptotics as $\abs{z}\to \infty$ while being analytic everywhere, including at $z = 0$.

This asymptotic analysis implies that  Hamiltonian in Eq.\ (\ref{hhh}) takes the form
\Beq
\hat{H} = \hat{H}_{a} + \hat{H}',
\Eeq
with
\Beq
\hat{H}' = \frac{Z}{\sqrt{\rho^{2}+z^{2}}} -  Z\frac{\erf{(a_{0} z)}}{z}.
\Eeq
Note that $\hat{H}'$ does not depend on the magnetic field strength.
Then, we can set up the homogeneous Lippmann-Schwinger equation
\Beq
\ket{\Psi} = \hat{G}_{a}(E) \hat{H}' \ket{\Psi},
\label{lseq}
 \Eeq
 where $\hat{G}_{a}(E) = (E -\hat{H}_{a})^{-1}$ is the Green's operator of the asymptotic Hamiltonian.
 The solution of this equation provides us with the bound and resonant states of the quantum mechanical system with negative real
 and complex energies, respectively.

\section{Solution method}

In solving the  Lippmann-Schwinger equation we follow an idea which was developed before for solving the three-body Faddeev 
equations for charged particles \cite{PhysRevC.54.50,PhysRevC.55.1080,PhysRevA.63.062721,papp2002faddeev,papp2002electron}.
We approximate $\hat{H}'$ in a discrete Hilbert space basis 
\Beq
\hat{H}' \simeq \sum_{i,j}^{N} \ket{i} \underline{H}'_{ij}  \bra{j},
\Eeq
where $N$ is finite and $\underline{H}'_{ij}=\mel{i}{\hat{H}'}{j}$. Then, the Lippmann-Schwinger equation becomes
\Beq
\ket{\psi} = \sum_{i,j}^{N} \hat{G}_{a}(E)  \ket{i}  \underline{H}'_{ij} \braket{j}{\psi}.
\label{eq12}
\Eeq
We should notice that  $\ket{\psi}$ on the left-hand side of Eq.\ \eqref{eq12} depends on the 
overlap $\braket{j}{\psi}$ on the right-hand side with $j$ up to $N$ only.
Therefore, to determine the coefficients,  it is sufficient to act from the left by $\bra{k}$, with $k$ up to $N$ only
\Beq
\braket{k}{\psi} = \sum_{i,j}^{N} \mel{k}{ \hat{G}_{a}(E)}{i} \underline{H}'_{ij}  \braket{j}{\psi}, \quad k = 1 \ldots N.
\Eeq
This is a matrix equation for the vector $\underline{\psi} =\braket{i}{\psi}$
\Beq
\underline{\psi} = \underline{G}_{a}(E) \underline{H}' \underline{\psi},
\Eeq
with matrix $\underline{G}_{a} = \mel{i}{ \hat{G}_{a}}{j}$.  
This equation is a homogeneous algebraic equation
\Beq
( \underline{G}^{-1}_{a}(E) - \underline{H}' )  \underline{\psi} =0,
\Eeq
and the energy eigenvalues can be determined from the condition
\Beq
\det ( \underline{G}^{-1}_{a}(E) - \underline{H}' ) =0.
\Eeq

Certainly, the evaluation of the matrix elements  $\mel{i}{\hat{H}'}{j}$ between any kind of basis functions can be done,
at least numerically. The real challenge is the calculation of the matrix elements of the Green's operator. Therefore,
the basis should be chosen to facilitate this task. 

In calculating the Green's operator, the general idea is to find a basis such that  the operator
$\hat{J} = \lambda -\hat{h}$, where $\lambda$ is a complex number, 
is represented by an infinite symmetric tridiagonal, i.e.\ Jacobi matrix.
It has been shown in Refs.\ \cite{Konya:1997JMP,PhysRevC.61.034302,PRADemir2006,kelbert2007green} 
that the $N\times N$ matrix representation 
of $\hat{g}(\lambda)=(\lambda-\hat{h})^{-1}$ is given by
\Beq
\underline{g}(\lambda)=( \underline{J} - \delta_{i,N}\delta_{j,N} J_{N,N+1} C_{N+1} J_{N+1,N})^{-1},
\Eeq
where $C_{N+1}$ is a continued fraction
such that
\Beq
C_{N+1}^{-1}= J_{N+1,N+1} - J_{N+1,N+2} ( C_{N+2}^{-1} )^{-1} J_{N+2,N+1}.
\Eeq

We should notice that $\hat{h}_{\rho}$  is, in fact, a Hamiltonian of a 
two-dimensional harmonic oscillator. 
The matrix elements  of the Green's operator of the $d$-dimensional harmonic oscillator 
can de given in terms of a continued fraction on the $d$-dimensional 
harmonic-oscillator basis \cite{Konya:1997JMP}.  The two-dimensional harmonic-oscillator  basis in given by
\Beq
\begin{split}
\braket{\rho}{\nu;\omega_{\rho}} = & \left( \frac{2 \sqrt{b} \Gamma(\nu+1)}{\Gamma(\nu+2 l_{z}+1)} \right)^{1/2} \exp(-b \rho^{2}/2)  \\
& \times (b \rho^{2})^{l_{z}/2+1/4} L_{\nu }^{(l_{z})}(b\rho^{2}),
\end{split}
\Eeq
where $L^{(\alpha)}_{\nu}$ is an associated Laguerre polynomial,
and $b$ is  related to the frequency parameter of the harmonic oscillator $b = m\omega_{\rho}/\hbar$.
The tridiagonal  matrix elements of  $J^{(\rho)} =   \lambda - \hat{h}_{\rho}$
 are given by 
\Beq
\begin{split}
&J^{(\rho)}_{\nu \nu'} =  \mel{\nu,\omega_{\rho}}{\lambda - \hat{h}_{\rho}}{\nu',\omega_{\rho}}    \\
&= \begin{cases} [ \lambda- \hbar (\omega_{h}^{2}+\omega_{\rho}^{2})/(2\omega_{\rho}) \cdot (2 \nu+l_{z}+1) ] \delta_{\nu \nu'}, \\ 
 \hbar (\omega_{h}^{2} -\omega_{\rho}^{2})/(2\omega_{\rho})  \cdot \sqrt{ (\nu+1)(\nu+1+l_{z})} \delta_{\nu \nu' -1}, \\ 
\hbar (\omega_{h}^{2} - \omega_{\rho}^{2})/(2\omega_{\rho})  \cdot \sqrt{\nu(\nu+l_{z})} \delta_{\nu \nu'+1}.
\end{cases}
\end{split}
\Eeq

As far as the Green's operator of $\hat{h}_{z}$ is concerned, we need a basis that reflects the even and odd $z$-parity of the system.
Here, we take the one-dimensional harmonic oscillator basis with $\omega_{z}$ frequency 
\Beq
\braket{z}{n,\omega_{z}} =   \left(   \frac{a}{  \sqrt{\pi} 2^{n}n! } \right)^{1/2} \exp(- a^{2} z^{2}/2 ) H_{n}( az), 
\Eeq
where $a = \sqrt{m \omega_{z}/\hbar}$.
An easy calculation gives for the non-vanishing 
matrix elements 
\Beq
\begin{split}
& \mel{n,\omega_{z}}{ \lambda - \hat{p}_{z}^{2}/2}{n',\omega_{z}} \\
& = \begin{cases} [ \lambda- \hbar \omega_{z}/2 \cdot (n+1/2) ]\delta_{n n'}, \\ 
\hbar \omega_{z}/4 \cdot \sqrt{(n+1)(n+2)} \delta_{n n'-2}, \\ 
\hbar \omega_{z}/4  \cdot \sqrt{n(n-1)} \delta_{n n'+2}.
\end{cases}
\end{split}
\Eeq
This matrix is not tridiagonal. However, if we select even parity and odd parity basis by taking $n$ even and odd, respectively, the matrix 
becomes tridiagonal and the procedure 
devised above results in the even and odd parity free Green's matrix $\underline{g}_{0}^{(e/o)}$. 
The Green's operator $\hat{g}_{z}(\lambda)=(\lambda -\hat{h}_{z})^{-1}$ satisfies the Lippmann-Schwinger equation
\Beq
\hat{g}_{z} = \hat{g}_{0} +  \hat{g}_{0} \hat{v}_{z} \hat{g}_{z},
\Eeq
which, with a discrete Hilbert space representation of $\hat{v}_{z}$, results in the 
even or odd $z$-parity Green's matrix 
\Beq
\underline{g}_{z}^{(e/o)} = ( (\underline{g}_{0}^{(e/o)} )^{-1}- \underline{v}_{z}^{e/o} )^{-1},
\Eeq
where  $\underline{v}_{z}^{e/o}  = \mel{n,\omega_{z}}{v(z)  }{n',\omega_{z}}$,
with $n$ and $n'$ being even or odd.

The Green's operator $\hat{G}_{a}(E)=(E-\hat{h}_{\rho}-\hat{h}_{z})^{-1}$ is the resolvent operator of two commuting Hamiltonians. 
Consequently, it can be expressed as a convolution integral 
\Beq
\hat{G}_{a}(E)  = \frac{1}{2 \pi i} \oint_{C} \dd{\lambda} \hat{g}_{z}(E-\lambda) \hat{g}_{\rho}(\lambda),
\label{convint}
\Eeq  
where the contour should be taken counterclockwise such that it goes around the spectrum of $\hat{h}_{\rho}$ 
without penetrating into the spectrum of 
$\hat{h}_{z}$ (see Refs.\ \cite{bianchi1964convolution} and 
\cite{PhysRevC.54.50,PhysRevC.55.1080,PhysRevA.63.062721,papp2002faddeev,papp2002electron}). 
We should note that the role of $\hat{g}_{\rho}$ and $\hat{g}_{z}$ are interchangeable.

\begin{figure}[!ht]
\centering
\includegraphics[width=8.4cm]{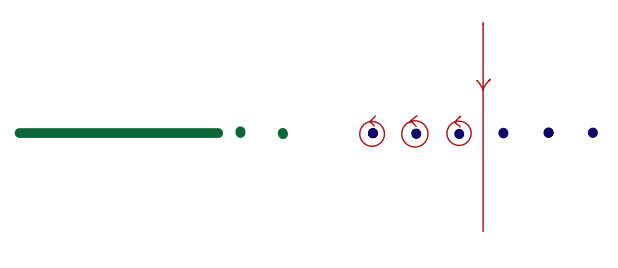}
\caption{Analytic structure of the integrand in Eq.\ (\ref{convint}). The poles of $\hat{g}_{\rho}$ are enclosed  such that
$\hat{g}_{z}$ is analytic in the enclosed region. The vertical line with a contour that encloses at infinity incorporates all the poles of $\hat{g}_{\rho}$. }
\label{fig1}
\end{figure}

With the adopted basis functions in $\rho$ and $z$, the matrix elements 
$\underline{g}_{\rho} = \mel{\nu , \omega_{\rho}}{\hat{g}_{\rho}}{\nu',\omega_{\rho}}$ and
$\underline{g}_{z} = \mel{n,\omega_{z}}{\hat{g}_{z}}{n';\omega_{z}}$ can be evaluated analytically on the whole complex energy plane. 
The energy parameter $E$ in Eq.\ (\ref{convint}) is such that  with a careful choice of the parameter $a_{0}$ in Eq.\ \eqref{erfpot}
we can achieve that  the two spectra are well separated and the
conditions for the contour integral is satisfied. Fig \ref{fig1} shows the analytic structure of the integrand with 
a possible contour in the complex $\lambda$ plane. The lowest lying poles of $\hat{g}_{\rho}$ are enclosed separately and 
 an integration path along an imaginary line together with a closing semi-circle at infinity, which vanishes anyway, incorporate the rest. 
 This path guaranties that
$\underline{g}_{z}$ is analytic on the encircled domain. 
  
The matrix elements of $\hat{H}'$ becomes a double integral 
$\underline{H}' = \mel{n, \omega_{z}; \nu, \omega_{\rho}}{ \hat{H}' }{n', \omega_{z}; \nu', \omega_{\rho}} $,
which can be evaluated numerically by using Gaussian integration rules. 
We choose $\omega_{z}$ and $\omega_{\rho}$ such that $a=\sqrt{b}$.
This way the basis functions in $z$ and $\rho$ have the same asymptotic behavior.

\section{Results}

\begin{figure}[!ht]
\centering
\includegraphics[width=8.5cm]{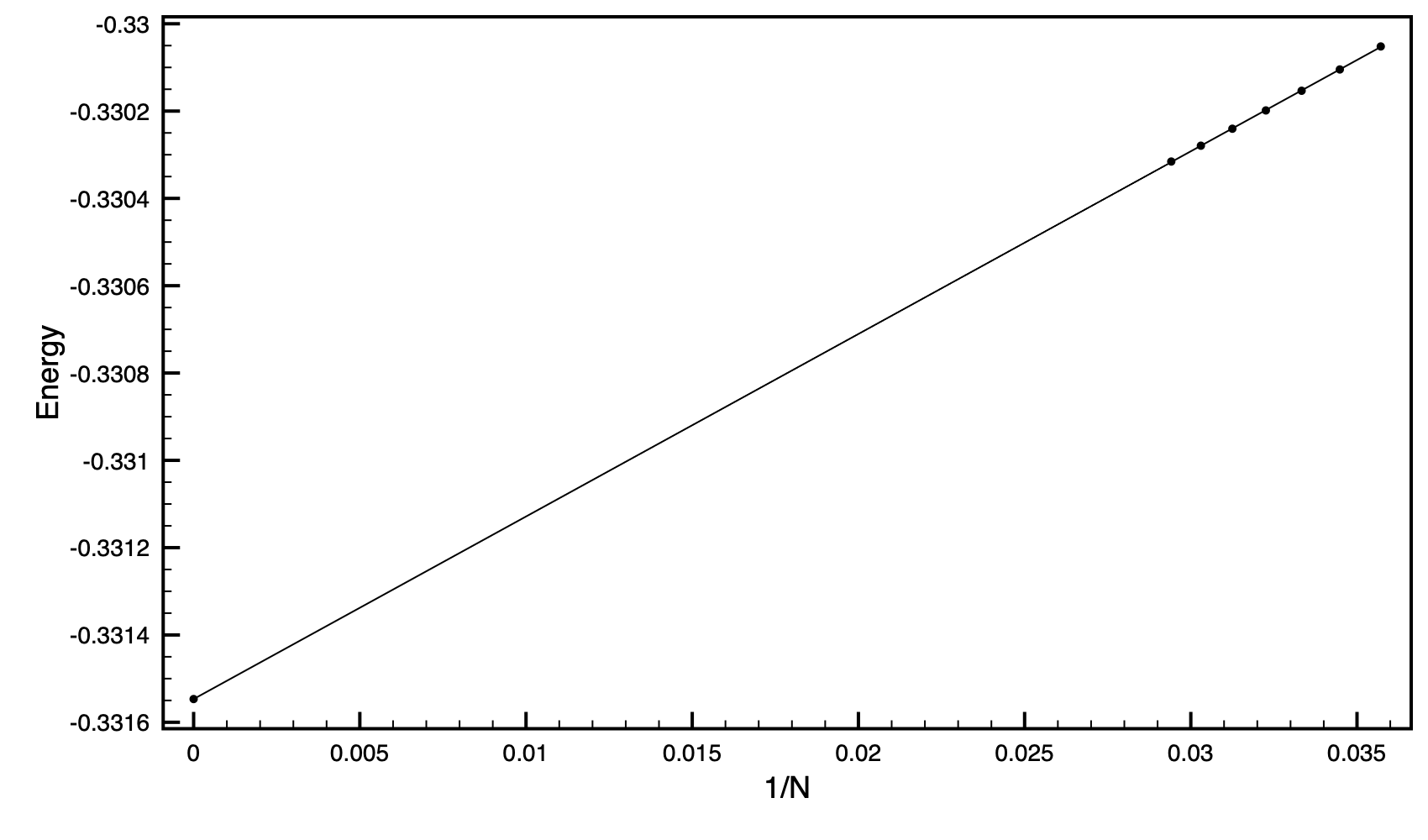}
\caption{Convergence of energy of the $l_z=0$, $\pi_z=1$ state with ${\cal H}=1$ as a function of $1/N$. }
\label{fig2}
\end{figure}

For a numerical illustration of our method, in Table \ref{tab1}, we show the lowest states 
with $\abs{l_{z}} \le 2$ and $\pi_{z} =\pm 1$. 
As a parameter for the HO basis 
in the coordinate $\rho$  we took $b=1.1$ and we calculated eigenstates with basis size from
$N=28$ to $N=34$ in $\rho$ and $z$. 
We observe an almost constant, slightly steady linear behavior for the energy  in $N$, 
which we could easily extrapolate to $N=\infty$. 
Fig \ref{fig2} shows the dependence of the energy in $1/N$, and its approximation to $1/\infty$. 
The method can provide about five digits accuracy, 
for all of the considered states and over the whole range of $\cal H$ values.
\begin{table}
\caption{
Energy eigenvalues of the Hamiltonian in Eq.\ (\ref{hhh}) with quantum numbers $l_{z}^{(\pi_{z})}$ for various values  of $\cal H$. }
\begin{center}
\begin{tabular}{|r|c|c|c|c| c | c| }
\hline
$\cal H$ & $0^{(+)}$ & $0^{(-)} $ &  $1^{(+)}$   &  $1^{(-)}$   &  $2^{(+)}$   &  $2^{(-)}$  \\
\hline 
 0.1 &  -0.49780  & -0.11271  & -0.10058  &    -0.00821 & 0.02526  &  0.07283  \\  
 0.2 &  -0.49067  & -0.08537  & -0.05054   &   0.06666 & 0.11872 &  0.18691   \\  
  0.3 &  -0.47949  & -0.05162  & 0.01091  &   0.15075  & 0.23598 &  0.32033   \\  
 0.4 &  -0.46493   & -0.01434  & 0.07864   & 0.23836  & 0.35900 &  0.45683  \\  
 0.5 &  -0.44755   & 0.02518   & 0.15052  &  0.32813  &  0.48560 &  0.59512  \\  
  0.6 & -0.42782   & 0.06628   & 0.22537   & 0.41939   & 0.61474 & 0.73464  \\  
 0.7 &  -0.40610   &0.10854   &  0.30249   & 0.51174  & 0.74580  &  0.87509  \\  
 0.8 &  -0.38268   & 0.15168  &  0.38141  &  0.60495 & 0.87836 &  1.01625  \\ 
 0.9 &  -0.35779   & 0.19554  &  0.46179  &  0.69882  & 1.01214 &  1.15800 \\ 
 1.0 &   -0.33161   &0.23998   & 0.54340   &  0.79325  &1.14695  & 1.30022  \\ 
  1.2 &  -0.27597    & 0.33022  & 0.73673  & 0.98340 &  1.41904 & 1.58583  \\ 
1.4 &   -0.21669   & 0.42186  &  0.90782 &  1.17490  & 1.69376 & 1.87266  \\ 
1.6&   -0.15445   & 0.51456  &  1.08128  & 1.36742   & 1.97054 & 2.16042  \\ 
 1.8 &  -0.08972    & 0.60809  &  1.25666 & 1.56074 & 2.24899 &  2.44895  \\ 
 2.0 &   -0.02625   & 0.70181  & 1.43364 & 1.75470   & 2.52883 & 2.73810  \\ 
  3.0 & 0.33458     & 1.17996   & 2.33564 & 2.73111   &3.94285  & 4.19048  \\
 4.0 &  0.71412    & 1.66425  &  3.25621 & 3.71418   & 5.37299 & 5.65005  \\ 
 5.0 &  1.11816    &2.15238   &  4.18837&  4.70104  & 6.81319 &  7.11417  \\ 
 6.0 &   1.53000   & 2.64283  &  5.12859 & 5.69036 & 8.26040 &  8.58150 \\ 
 7.0 &  1.95064    & 3.13493  &  6.07482 & 6.68140  & 9.71284 &  10.0513  \\ 
  8.0 &   2.37815   & 3.62822  & 7.02573 & 7.67371  & 11.1694 & 11.5231   \\ 
 9.0 &  2.81121    & 4.12242    & 7.98041&  8.66699   & 12.6292 & 12.9964  \\ 
 10.0&  3.24885    & 4.61734  &  8.93819 & 9.66104   & 14.0917 & 14.4712   \\           
\hline
\end{tabular}
\label{tab1}
\end{center}
\end{table}

\section{Summary and conclusions}

In this work we propose a novel method for solving the problem of a hydrogen atom in a strong magnetic field.
We define the asymptotic Hamiltonian as a sum of a two-dimensional harmonic oscillator in the $\rho$ and 
a one-dimensional Coulomb-like Hamiltonian in the $z$ coordinate. With the separation of the Hamiltonian into asymptotically 
relevant and asymptotically irrelevant parts, we can set up a Lippmann-Schwinger equation with the Green's operator containing
the asymptotic Hamiltonian and the source term consisting of the asymptotically irrelevant parts. 
The Hilbert space basis is defined as the direct product of bases in coordinates $\rho$ and $z$, and a finite basis representation
of the source term turns the Lippmann-Schwinger equation into a matrix equation.
Those bases in $\rho$ and $z$  are chosen to facilitate the evaluation of the respective Green's operators, from which 
the Green's operator of asymptotic Hamiltonian can be calculated by a complex contour integral. 
The method seems rather robust. We could achieve high accuracy over a large range of $\cal H$ values using only one choice 
of parameter $b$. 

The problem of a hydrogen atom in a strong magnetic field is an old classic problem of quantum mechanics. 
New approaches to old problems not only corroborate old result, but may open new possibilities as well. The proposed
method solves an integral equation, where the boundary conditions are incorporated in the resolvent operator. 
We are sure that this approach opens new ways of attacking problems like this.

\bibliographystyle{apsrev4-1} 
\bibliography{fv00}

\begin{thebibliography}{12}%
\makeatletter
\providecommand \@ifxundefined [1]{%
 \@ifx{#1\undefined}
}%
\providecommand \@ifnum [1]{%
 \ifnum #1\expandafter \@firstoftwo
 \else \expandafter \@secondoftwo
 \fi
}%
\providecommand \@ifx [1]{%
 \ifx #1\expandafter \@firstoftwo
 \else \expandafter \@secondoftwo
 \fi
}%
\providecommand \natexlab [1]{#1}%
\providecommand \enquote  [1]{``#1''}%
\providecommand \bibnamefont  [1]{#1}%
\providecommand \bibfnamefont [1]{#1}%
\providecommand \citenamefont [1]{#1}%
\providecommand \href@noop [0]{\@secondoftwo}%
\providecommand \href [0]{\begingroup \@sanitize@url \@href}%
\providecommand \@href[1]{\@@startlink{#1}\@@href}%
\providecommand \@@href[1]{\endgroup#1\@@endlink}%
\providecommand \@sanitize@url [0]{\catcode `\\12\catcode `\$12\catcode
  `\&12\catcode `\#12\catcode `\^12\catcode `\_12\catcode `\%12\relax}%
\providecommand \@@startlink[1]{}%
\providecommand \@@endlink[0]{}%
\providecommand \url  [0]{\begingroup\@sanitize@url \@url }%
\providecommand \@url [1]{\endgroup\@href {#1}{\urlprefix }}%
\providecommand \urlprefix  [0]{URL }%
\providecommand \Eprint [0]{\href }%
\providecommand \doibase [0]{http://dx.doi.org/}%
\providecommand \selectlanguage [0]{\@gobble}%
\providecommand \bibinfo  [0]{\@secondoftwo}%
\providecommand \bibfield  [0]{\@secondoftwo}%
\providecommand \translation [1]{[#1]}%
\providecommand \BibitemOpen [0]{}%
\providecommand \bibitemStop [0]{}%
\providecommand \bibitemNoStop [0]{.\EOS\space}%
\providecommand \EOS [0]{\spacefactor3000\relax}%
\providecommand \BibitemShut  [1]{\csname bibitem#1\endcsname}%
\let\auto@bib@innerbib\@empty
\bibitem [{\citenamefont {Rau}(2012)}]{rau2012topics}%
  \BibitemOpen
  \bibfield  {author} {\bibinfo {author} {\bibfnamefont {A.}~\bibnamefont
  {Rau}},\ }\href@noop {} {\bibfield  {journal} {\bibinfo  {journal} {American
  Journal of Physics}\ }\textbf {\bibinfo {volume} {80}},\ \bibinfo {pages}
  {406} (\bibinfo {year} {2012})}\BibitemShut {NoStop}%
\bibitem [{\citenamefont {Elliott}\ and\ \citenamefont
  {Loudon}(1960)}]{elliott1960theory}%
  \BibitemOpen
  \bibfield  {author} {\bibinfo {author} {\bibfnamefont {R.}~\bibnamefont
  {Elliott}}\ and\ \bibinfo {author} {\bibfnamefont {R.}~\bibnamefont
  {Loudon}},\ }\href@noop {} {\bibfield  {journal} {\bibinfo  {journal}
  {Journal of Physics and Chemistry of Solids}\ }\textbf {\bibinfo {volume}
  {15}},\ \bibinfo {pages} {196} (\bibinfo {year} {1960})}\BibitemShut
  {NoStop}%
\bibitem [{\citenamefont {Papp}\ and\ \citenamefont
  {Plessas}(1996)}]{PhysRevC.54.50}%
  \BibitemOpen
  \bibfield  {author} {\bibinfo {author} {\bibfnamefont {Z.}~\bibnamefont
  {Papp}}\ and\ \bibinfo {author} {\bibfnamefont {W.}~\bibnamefont {Plessas}},\
  }\href@noop {} {\bibfield  {journal} {\bibinfo  {journal} {Phys. Rev. C}\
  }\textbf {\bibinfo {volume} {54}},\ \bibinfo {pages} {50} (\bibinfo {year}
  {1996})}\BibitemShut {NoStop}%
\bibitem [{\citenamefont {Papp}(1997)}]{PhysRevC.55.1080}%
  \BibitemOpen
  \bibfield  {author} {\bibinfo {author} {\bibfnamefont {Z.}~\bibnamefont
  {Papp}},\ }\href@noop {} {\bibfield  {journal} {\bibinfo  {journal} {Phys.
  Rev. C}\ }\textbf {\bibinfo {volume} {55}},\ \bibinfo {pages} {1080}
  (\bibinfo {year} {1997})}\BibitemShut {NoStop}%
\bibitem [{\citenamefont {Papp}\ \emph {et~al.}(2001)\citenamefont {Papp},
  \citenamefont {Hu}, \citenamefont {Hlousek}, \citenamefont {K\'onya},\ and\
  \citenamefont {Yakovlev}}]{PhysRevA.63.062721}%
  \BibitemOpen
  \bibfield  {author} {\bibinfo {author} {\bibfnamefont {Z.}~\bibnamefont
  {Papp}}, \bibinfo {author} {\bibfnamefont {C.-Y.}\ \bibnamefont {Hu}},
  \bibinfo {author} {\bibfnamefont {Z.~T.}\ \bibnamefont {Hlousek}}, \bibinfo
  {author} {\bibfnamefont {B.}~\bibnamefont {K\'onya}}, \ and\ \bibinfo
  {author} {\bibfnamefont {S.}~\bibnamefont {Yakovlev}},\ }\href@noop {}
  {\bibfield  {journal} {\bibinfo  {journal} {Phys. Rev. A}\ }\textbf {\bibinfo
  {volume} {63}},\ \bibinfo {pages} {062721} (\bibinfo {year}
  {2001})}\BibitemShut {NoStop}%
\bibitem [{\citenamefont {Papp}\ \emph {et~al.}(2002)\citenamefont {Papp},
  \citenamefont {Darai}, \citenamefont {Nishimura}, \citenamefont {Hlousek},
  \citenamefont {Hu},\ and\ \citenamefont {Yakovlev}}]{papp2002faddeev}%
  \BibitemOpen
  \bibfield  {author} {\bibinfo {author} {\bibfnamefont {Z.}~\bibnamefont
  {Papp}}, \bibinfo {author} {\bibfnamefont {J.}~\bibnamefont {Darai}},
  \bibinfo {author} {\bibfnamefont {A.}~\bibnamefont {Nishimura}}, \bibinfo
  {author} {\bibfnamefont {Z.}~\bibnamefont {Hlousek}}, \bibinfo {author}
  {\bibfnamefont {C.-Y.}\ \bibnamefont {Hu}}, \ and\ \bibinfo {author}
  {\bibfnamefont {S.}~\bibnamefont {Yakovlev}},\ }\href@noop {} {\bibfield
  {journal} {\bibinfo  {journal} {Physics Letters A}\ }\textbf {\bibinfo
  {volume} {304}},\ \bibinfo {pages} {36} (\bibinfo {year} {2002})}\BibitemShut
  {NoStop}%
\bibitem [{\citenamefont {Papp}\ and\ \citenamefont
  {Hu}(2002)}]{papp2002electron}%
  \BibitemOpen
  \bibfield  {author} {\bibinfo {author} {\bibfnamefont {Z.}~\bibnamefont
  {Papp}}\ and\ \bibinfo {author} {\bibfnamefont {C.-Y.}\ \bibnamefont {Hu}},\
  }\href@noop {} {\bibfield  {journal} {\bibinfo  {journal} {Physical Review
  A}\ }\textbf {\bibinfo {volume} {66}},\ \bibinfo {pages} {052714} (\bibinfo
  {year} {2002})}\BibitemShut {NoStop}%
\bibitem [{\citenamefont {K{\'o}nya}\ \emph {et~al.}(1997)\citenamefont
  {K{\'o}nya}, \citenamefont {L{\'e}vai},\ and\ \citenamefont
  {Papp}}]{Konya:1997JMP}%
  \BibitemOpen
  \bibfield  {author} {\bibinfo {author} {\bibfnamefont {B.}~\bibnamefont
  {K{\'o}nya}}, \bibinfo {author} {\bibfnamefont {G.}~\bibnamefont
  {L{\'e}vai}}, \ and\ \bibinfo {author} {\bibfnamefont {Z.}~\bibnamefont
  {Papp}},\ }\href@noop {} {\bibfield  {journal} {\bibinfo  {journal}
  {J.Math.Phys.}\ }\textbf {\bibinfo {volume} {38}},\ \bibinfo {pages} {4832}
  (\bibinfo {year} {1997})}\BibitemShut {NoStop}%
\bibitem [{\citenamefont {K{\'o}nya}\ \emph {et~al.}(2000)\citenamefont
  {K{\'o}nya}, \citenamefont {L{\'e}vai},\ and\ \citenamefont
  {Papp}}]{PhysRevC.61.034302}%
  \BibitemOpen
  \bibfield  {author} {\bibinfo {author} {\bibfnamefont {B.}~\bibnamefont
  {K{\'o}nya}}, \bibinfo {author} {\bibfnamefont {G.}~\bibnamefont
  {L{\'e}vai}}, \ and\ \bibinfo {author} {\bibfnamefont {Z.}~\bibnamefont
  {Papp}},\ }\href@noop {} {\bibfield  {journal} {\bibinfo  {journal} {Phys.
  Rev. C}\ }\textbf {\bibinfo {volume} {61}},\ \bibinfo {pages} {034302}
  (\bibinfo {year} {2000})}\BibitemShut {NoStop}%
\bibitem [{\citenamefont {Demir}\ \emph {et~al.}(2006)\citenamefont {Demir},
  \citenamefont {Hlousek},\ and\ \citenamefont {Papp}}]{PRADemir2006}%
  \BibitemOpen
  \bibfield  {author} {\bibinfo {author} {\bibfnamefont {F.}~\bibnamefont
  {Demir}}, \bibinfo {author} {\bibfnamefont {Z.~T.}\ \bibnamefont {Hlousek}},
  \ and\ \bibinfo {author} {\bibfnamefont {Z.}~\bibnamefont {Papp}},\
  }\href@noop {} {\bibfield  {journal} {\bibinfo  {journal} {Phys. Rev. A}\
  }\textbf {\bibinfo {volume} {74}},\ \bibinfo {pages} {014701} (\bibinfo
  {year} {2006})}\BibitemShut {NoStop}%
\bibitem [{\citenamefont {Kelbert}\ \emph {et~al.}(2007)\citenamefont
  {Kelbert}, \citenamefont {Hyder}, \citenamefont {Demir}, \citenamefont
  {Hlousek},\ and\ \citenamefont {Papp}}]{kelbert2007green}%
  \BibitemOpen
  \bibfield  {author} {\bibinfo {author} {\bibfnamefont {E.}~\bibnamefont
  {Kelbert}}, \bibinfo {author} {\bibfnamefont {A.}~\bibnamefont {Hyder}},
  \bibinfo {author} {\bibfnamefont {F.}~\bibnamefont {Demir}}, \bibinfo
  {author} {\bibfnamefont {Z.~T.}\ \bibnamefont {Hlousek}}, \ and\ \bibinfo
  {author} {\bibfnamefont {Z.}~\bibnamefont {Papp}},\ }\href@noop {} {\bibfield
   {journal} {\bibinfo  {journal} {Journal of Physics A: Mathematical and
  Theoretical}\ }\textbf {\bibinfo {volume} {40}},\ \bibinfo {pages} {7721}
  (\bibinfo {year} {2007})}\BibitemShut {NoStop}%
\bibitem [{\citenamefont {Bianchi}\ and\ \citenamefont
  {Favella}(1964)}]{bianchi1964convolution}%
  \BibitemOpen
  \bibfield  {author} {\bibinfo {author} {\bibfnamefont {L.}~\bibnamefont
  {Bianchi}}\ and\ \bibinfo {author} {\bibfnamefont {L.}~\bibnamefont
  {Favella}},\ }\href@noop {} {\bibfield  {journal} {\bibinfo  {journal} {Il
  Nuovo Cimento (1955-1965)}\ }\textbf {\bibinfo {volume} {34}},\ \bibinfo
  {pages} {1825} (\bibinfo {year} {1964})}\BibitemShut {NoStop}%
\end{thebibliography}%

\end{document}